# Working paper

# Boucle rétroactive entre la volatilité des flux de capitaux et la stabilité financière : résultat pour la République démocratique du Congo

Christian Pinshi Paula[1]

**RÉSUME**

Le système financier étant le lieu de rencontre des flux de capitaux (égalité entre épargne et investissement), une volatilité des flux de capitaux peut détruire le bon fonctionnement et la robustesse du système financier, c'est-à-dire saper la stabilité financière. De même un système financier faible, peu réglementé et mal géré peut exacerber la volatilité des flux de capitaux et in fine saper la stabilité financière. Cette étude examine la boucle rétroactive entre la volatilité des flux de capitaux et la stabilité financière en République démocratique du Congo (RDC), et évalue les contributions des politiques macroéconomique et macroprudentielle dans l'atténuation des effets de la volatilité des flux de capitaux sur la stabilité financière et dans la prévention de l'instabilité financière. L'estimation du modèle de régression dynamique à la Feldstein-Horioka nous a montré que le système financier est largement alimenté et financé par les flux de capitaux internationaux. Ceci implique que l'économie congolaise est mobile financièrement, ceci peut être un danger pour la stabilité financière. L'étude économétrique dynamique de la taille absolue du système financier, nous stipule que le système financier a un poids systémique sur l'économie réelle. D'où un choc du système financier pourrait avoir des effets dévastateurs sur l'économie congolaise. Nous estimons un modèle vectoriel autorégressif pour vérifier la causalité bidirectionnelle et les impacts des politiques macroéconomique et macroprudentielle. Eu égards aux résultats, il s'est avéré d'une part, qu'il y a une boucle rétroactive entre la volatilité des flux de capitaux et la stabilité financière. Et d'autre part, que les politiques macroéconomique et macroprudentielle ne peuvent pas atténuer la volatilité des flux de capitaux et prévenir une instabilité financière. Il s'avère que l'approche macroprudentielle a donné un résultat meilleur que la politique monétaire. La mise en œuvre d'un cadre macroprudentielle par la Banque centrale du Congo sera bénéfique dans l'atteinte de la stabilité financière et dans l'atténuation de la volatilité des flux de capitaux.

Mots-clés: volatilité des flux des capitaux, stabilité financière, politique macroéconomique et macroprudentielle.

Code JEL: F32, E37, E58, G13, G18

---

[1] Chercheur en macroéconomie et en système financier. Phone: +243 (0) 81 570 25 30.
E-mail: elijahpinshicolorado@gmail.com.

# Working paper

# Feedback effect between Volatility of capital flows and financial stability: evidence from Democratic Republic of Congo

## Christian Pinshi Paula

**ABSTRACT**

Financial system being the place of metting capital flows (equality between saving and investment), a volatility of capital flows can destroy the robustness and good working of financial system, it means subvert financial stability. The same a weak financial system, few regulated and bad manage can exacerbate volatility of capital flows and finely undermine financial stability. The present study provides evidence on feedback effect between volatility of capital flows and financial stability in Democratic republic of Congo (DRC), and estimate the contributions of macroeconomic and macroprudential policies in the attenuation volatility of capital flows effects on financial stability and in the prevention of instability financial. Assessment dynamic regression model a la Feldstein-Horioka we showed that financial system is widely supplied and financed by internationals capital flows. This implicate Congolese economy is financially mobile, that can be dangerous for financial stability. The study dynamic econometric of financial system's absolute size, we stipulate financial system has a systemic weight on real economy. Hence a shock of financial system could have devastating effects on Congolese economy. We estimate a vector autoregressive (VAR) model for prove the bilateral causality and impacts of macroeconomic and macroprudential policies. With regard to results, it proved on the one there is a feedback effect between volatility of capital flows and financial stability, on the other hand macroeconomic and macroprudential policies can't attenuate volatility of capital flows and prevent instability financial. It prove macroprudential approach is given a better result than monetary policy. The implementation of framework macroprudential by Central Bank of Congo will be beneficial in the realization of financial stability and attenuation volatility of capital flows.

# 1 | INTRODUCTION

Au cours de quatre dernières décennies, les flux de capitaux se sont intensifiés, impulsés par la libéralisation financière. Les flux de capitaux peuvent contribuer à augmenter la production, soutenir la demande interne et favoriser la croissance et/ou le revenu par tête et l'emploi. Outre ces gains, les flux de capitaux peuvent aussi se traduire par l'approfondissement du système financier. Les facteurs attractifs de ces flux sont liés à l'amélioration de fondamentaux économiques, tels que la stabilité macroéconomique et l'ouverture du compte financier, mais également à la qualité de l'environnement politique. La capacité d'un pays en développement à saisir les avantages de ces flux de capitaux ou sa fragilité à la volatilité de ces derniers peuvent être considérablement affectées par la qualité de son cadre macroéconomique et la force de son système financier.

Le système financier étant le lieu de rencontre des flux de capitaux (égalité entre épargne et investissement), une volatilité des flux de capitaux peut détruire le bon fonctionnement et la robustesse du système financier, c'est-à-dire saper la stabilité financière. De même un système financier faible, peu réglementé et mal gérer peut exacerber la volatilité des flux de capitaux et in fine saper la stabilité financière.

Au cours de la décennie 1990, les flux de capitaux ont été principalement dirigés vers les pays émergents de l'Asie, notamment la Malaisie et le Thaïlande, entrainant une forte croissance économique. Cependant, il convient de relever que la volatilité des flux de capitaux peut également s'accompagné des déséquilibres macroéconomiques (inflation élevé, déficit du compte courant, déficit budgétaire,…) et des perturbations de la stabilité du système financier : des crises bancaire, de change, d'endettement puis financière atteignant la sphère réelle de l'économie (production, chômage,…). A cet effet, ces flux peuvent être à l'origine de grandes catastrophes. Cette volatilité est en partie attribuée au fait que les déficits et excédents des balances courantes des pays se sont accentués.

Le processus d'intégration financière des pays de l'Asie de l'Est a fortement modifié le comportement des banques et des investisseurs internationaux, donnant lieu à des vagues alternés d'entrées et de sorties de capitaux aux effets perturbateurs qui se sont traduits par la volatilité des flux de capitaux. Cette situation a conduit à la crise financière asiatique de 1997, masquant une épargne intérieure largement insuffisante. Le déclencheur immédiat de la crise financière en Asie a été la volatilité de flux de capitaux (à des entrées nettes de capitaux représentant 6% du PIB en 1995 dans les pays asiatiques touchés par la crise succédèrent des sorties nettes équivalant à 2% du PIB en 1997, puis 5% du PIB l'année suivante), mais la cause fondamentale en a été la faiblesse du système financier qui a laissé de nombreuses économies vulnérables au retrait soudain des capitaux. Cette crise a permis de mettre en évidence les faiblesses du système financier (Burton et al, 2006). Depuis la crise asiatique de 1997, la plupart de pays ont renforcé sensiblement leur système financier par des réformes. Il en est résulté un meilleur équilibre de la composition des flux de capitaux. Malgré ces progrès, le développement des liens financiers étroits avec les marchés financiers internationaux a contribué à rendre beaucoup de pays vulnérables aux perturbations externes qui se sont produites en 2008 (Suchanek et Vasishtha, 2010).

En 2001 la RDC libéralisa son compte de capital, enfin que les flux de capitaux internationaux puissent se diriger vers son économie et son secteur financier, ceci en contribuant à accroitre le développement financier. La littérature nous dit qu'une libéralisation financière contribue à l'approfondissement du système financier et au développement économique. Cependant la libéralisation financière peut fragiliser le système financier et accentuer la volatilité des flux de capitaux, ceci aura pour conséquence la contraction de l'économie et l'instabilité macroéconomico-financière. C'est ainsi que la libéralisation financière doit remplir certaines conditions d'ordre macroéconomique, financière et politique pour se protéger contre le sudden stop et les crises financières.

L'économie congolaise connait un déséquilibre financier, le niveau négatif du solde épargne-investissement caractérise les afflux massifs de capitaux vers la RDC (graphique 1.1). Bien que ces entrées de capitaux soient profitables à l'économie, elles créent des lourds déséquilibres financiers. Ces déséquilibres sont susceptibles de déstabiliser les flux internationaux de capitaux, à l'effet de se répercuter sur la stabilité du système financier et ainsi préparer le terrain pour une phase de stress financier[2].

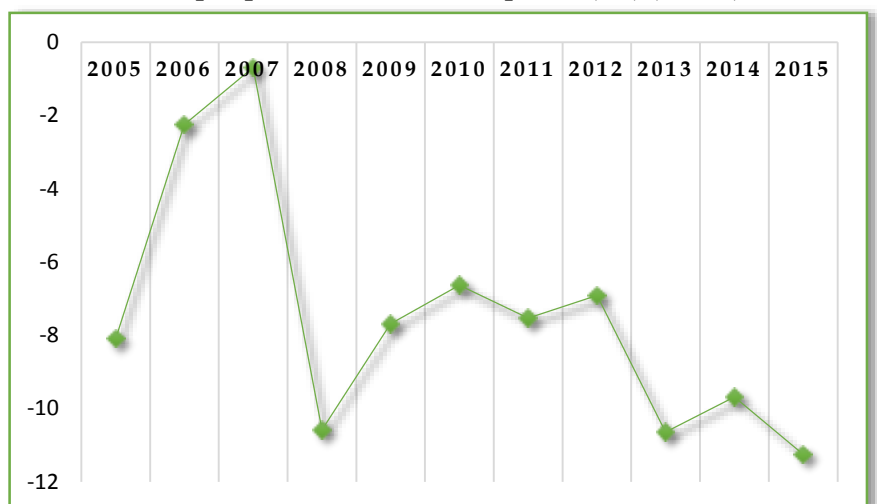

**Graphique 1.1 Flux net de capitaux (S-I) (% PIB)**

*Source : FMI, Perspective de l'économie mondiale (octobre, 2016)*

---

[2] Le stress financier est une perturbation déclenché par le comportement du système financier (faillite d'institutions financières, prise de risque, trouble mental du réseau financier,…) qui stimule la tension nerveuse du système financier (dysfonctionnement).
Un stress financier intense ou prolongé peut être source de divers troubles financiers et peut causer de perturbations majeures à l'économie réelle, c'est-à-dire une crise financière.

Eu égard à l'expérience vécue par les pays asiatiques avant, et pendant la crise de 1997-98, à la libéralisation financière, par la RDC depuis 2001 et à l'accumulation des déséquilibres internes et externes par le pays sur longue période, d'une part, l'on s'interroge sur : *l'existence d'une boucle rétroactive entre la volatilité des flux de capitaux et la stabilité financière en RDC ?* Car si ces flux de capitaux sont volatiles cela saperait le système financier et encore moins si le système financier est peu profond cela amplifierait le choc. La causalité unidirectionnelle des flux vers le système financier est considérée comme un danger pour la stabilité financière. Cependant si il y a boucle rétroactive c'est-à-dire une causalité bidirectionnelle entre la volatilité des flux de capitaux et la stabilité financière, ceci serait avantageux pour la stabilité économique et financière. Cette boucle serait une voie pour renforcer la régulation des politiques visant à atténuer la volatilité des flux de capitaux et traduire un processus de rattrapage et/ou de développement financier.

C'est dans le cadre d'une boucle rétroactive que l'on s'interroge d'autre part sur : *Quel rôle joue les politiques macroéconomique et prudentielle ou peuvent-elles jouer dans : l'atténuation des effets de la volatilité de flux des capitaux sur la stabilité financière ? Et dans la prévention des crises financière et la stabilité financière en RDC ?* Car un renforcement de la régulation financière pourrait être une condition pertinente pour prévenir les crises financières et réaliser la stabilité financière.

L'objectif principal de cette étude est d'analyser les interactions entre la volatilité des flux de capitaux et la stabilité financière en RDC, et son objectif secondaire est de proposer des politiques pour le renforcement de la gestion de la stabilité financière en RDC.

## 1|1 LA LITTÉRATURE

L'interaction entre la volatilité des flux de capitaux et la stabilité financière a donné lieu à une riche littérature théorique et empirique.

Lopez-Mejía (1999), relève que les entrées massives de capitaux peuvent provoquer un accroissement excessif de la demande globale (qui se traduirait par une surchauffe macroéconomique (pression inflationniste)), une appréciation du taux de change réel et une aggravation des déficits courants). Elles peuvent ainsi avoir des répercussions négatives sur le système financier (augmentation excessive du crédit bancaire, tensions macroéconomiques,…). Il souligne que les pays qui ont connu une expansion du crédit n'ont pas tous vu leur système financier s'affaiblir, et l'amplitude du cycle d'expansion et de repli a varié d'un pays à l'autre.

Burton et al. (2006) montrent que la volatilité des flux de capitaux déclenche l'instabilité financière, surtout dans un contexte de faiblesse du système financier. Il s'ensuit des retraits massifs des capitaux. D'où la causalité est bidirectionnelle.

Kose et al. (2007), démontrent théoriquement la boucle rétroactive entre la volatilité des flux de capitaux et la stabilité financière. Pour ces auteurs, plus le système financier est développé, plus les entrées de capitaux dopent la croissance et moins le pays est vulnérable aux crises, que ce soit de façon directe ou indirecte. L'essor du système financier a aussi un effet bénéfique sur la stabilité macroéconomique, qui se répercute elle-même sur le volume et la composition des flux de capitaux. Cependant, dans les pays en développement dont le système financier n'est pas assez dense, les reflux soudains de capitaux tendent à engendrer des cycles alternant expansion et repli ou à amplifier la volatilité. Ceci contribue au déclenchement d'une instabilité financière (c'est-à-dire une crise financière).

Bernanke (2011), pour retracer plus complètement l'incidence des flux de capitaux sur la stabilité financière, analyse théoriquement le comportement du solde épargne diminué de l'investissement (S – I). Selon lui, les flux de capitaux ont contribué au financement des booms des actifs financiers et ainsi préparé le terrain pour la phase de l'instabilité financière dans les pays où le système financier est peu profond et moins règlementé. Il montre que l'interaction entre volatilité des flux de capitaux et faiblesse du système financier produit des effets dévastateurs à l'économie.

Kaminsky et Reinhart (1999) étudient la relation entre les crises financières (bancaires, monétaire,…) et les flux de capitaux dans une économie libéralisée. En utilisant un modèle Probit en panel, elles identifient une croissance excessive du crédit comme le facteur majeur du déclenchement d'une crise financière. La libéralisation financière stimule les entrées de capitaux qui se traduisent par un excès de liquidité et peuvent ainsi conduire à un accroissement du crédit bancaire et de la monnaie en circulation. Lorsque ces entrées massives de capitaux dans l'économie sont intermédiées par un système financier sous développé et peu réglementé, elles entraînent une volatilité de la consommation et, par conséquent, des importations. Tandis que l'investissement reste faible, l'économie devient alors plus vulnérable aux chocs exogènes.

Sa (2006) ne trouve pas de lien de causalité univoque entre flux de capitaux et crédit bancaire au secteur privé. Il est donc difficile d'en tirer des conclusions générales pour la stabilité financière. Toutefois, il observe que dans certains pays, les afflux massifs de capitaux étrangers et l'expansion du crédit peuvent être liés à un mouvement plus sain de processus d'approfondissement financier. Dans d'autres pays, la co-existence de fortes entrées de capitaux et de booms du crédit peut générer des risques d'instabilité lorsque ces financements externes amènent des profonds déséquilibres macroéconomiques et financiers.

Diev et Pouvelle (2008) estiment un modèle économétrique qui montre une relation significative et négative entre la croissance des crédits et les flux de capitaux. Ils trouvent qu'une hausse d'un point de pourcentage du ratio de crédit sur PIB détériorerait le ratio compte courant sur PIB de 0,5 point de pourcentage. Ils montrent qu'une croissance excessive des crédits contribuerait à détériorer le compte courant et/ou le compte financier au-delà d'un niveau jugé soutenable et augmenterait la probabilité d'une crise de change.

Puis ils mettent en œuvre des tests de causalité pour évaluer la nature de la croissance des crédits. Ils concluent que lorsque la causalité détectée va dans le sens de la croissance des crédits vers la demande interne et non l'inverse, ceci pourrait être interprété comme un risque potentiel pour la stabilité financière dans la mesure où la forte croissance des crédits entraîne une surchauffe allant au-delà des évolutions liées au simple rattrapage (processus de développement financier).

Furceri et al. (2011) étudient la relation entre les entrées des capitaux internationaux et la probabilité de déclenchement des crises financières. Pour ce faire, les auteurs estiment un modèle de régression logistique en données de panel sur un échantillon de 112 pays développés et émergents entre 1970 et 2007. Ils trouvent qu'un épisode d'afflux de capitaux internationaux augmente significativement la probabilité des crises financières dans les deux années qui suivent. Ils montrent aussi que la probabilité de déclenchement des crises est plus importante si les entrées de capitaux sont composées principalement de flux de dettes à court terme.

Pour Reinhart et Reinhart (2008), la probabilité d'une instabilité financière conditionnelle à un bonanza de capitaux est plus élevée que la probabilité non conditionnelle. Leur étude montre qu'une instabilité financière est beaucoup plus causée par un bonanza de capitaux.

Reinhart et Rogoff (2010), en calculant la corrélation entre mobilité des capitaux et instabilité financière entre 1800 et 2000, ont montré que les périodes de forte mobilité des capitaux ont de manière répétée causé des crises financières.

Janus et Riera-Crichton (2016), étudient la relation entre les crises financières et les flux de capitaux. Ils montrent que les crises bancaires associées au renversement des capitaux augmentent le risque de change et le sudden stop.

Magud et al. (2014), ont trouvé que dans un régime de flexibilité du taux de change, une politique macroprudentielle, permet aux pays d'atténuer et de lisser la volatilité des flux de capitaux.

Lambert et al. (2011), estiment un modèle Panel VAR pour saisir les effets spillover des taxes sur les capitaux. Ils concluent, que la hausse de la taxe imposée par le Brésil sur les flux de capitaux a entrainé une augmentation des flux de capitaux vers les autres pays. Ceci a coûté une perte des capitaux pour les Brésil et un gain important pour les autres pays voisins.

Glockert et Towbin (2012), analysent l'impact de la politique monétaire sur les flux de capitaux tout en poursuivant deux objectifs, à savoir la stabilité des prix et la stabilité financière. Par le truchement du modèle VAR, ils trouvent que l'impact du taux directeur sur les flux de capitaux est important si l'objectif poursuivi est la stabilité des prix, et moins important dans le cas de l'objectif de la stabilité financière. Contrairement au taux directeur, ils trouvent qu'un choc ou une politique des réserves obligatoires influent sur les flux de capitaux si l'objectif poursuivis est la stabilité financière.

Bussiere et al. (2015), estiment en panel la relation entre le niveau de réserves de change et le contrôle des capitaux. Les résultats indiquent que le niveau de réserves joue un rôle important : les pays ayant un niveau de réserves élevé (en proportion de leur dette à court terme) ont relativement moins souffert de la crise, particulièrement quand leur compte de capital est moins ouvert. Cela suggère une certaine complémentarité entre accumulation de réserves et contrôles des capitaux.

Clerc et al. (2010) suggèrent que la mise en place d'une politique macroprudentielle, sous la forme d'une taxe prudentielle des flux de capitaux ou de contrôle des capitaux, présente un intérêt limité, relativement à une intervention publique ex post en temps de crise

Un bilan général de la littérature évoquée ci-dessus révèle que certains auteurs prônent que la volatilité des flux de capitaux sape la stabilité financière et a des effets dévastateurs sur l'économie réelle. Cependant, beaucoup ont montré également qu'un système financier moins solide, sous développé et peu réglementé, peut causer la volatilité de flux de capitaux. Il s'en découle que l'effet causal n'est essentiellement pas univoque mais rétroactive c'est-à-dire l'influence est dans le deux sens.

Dans ce travail nous allons procéder à une analyse théorique et empirique d'une part de la relation que peut exercer la volatilité des flux de capitaux (stabilité financière) sur la stabilité financière (la volatilité des flux de capitaux) c'est-à-dire la manière dont elles s'influent mutuellement , et d'autre part, de la capacité des politiques macroéconomiques et macroprudentielle à atténuer la volatilité des flux de capitaux et à prévenir le risque de crise dans la but de préserver la stabilité financière.
Cependant, ce papier est organisé comme suit. La première section présente un cadre analytique de l'interaction entre la stabilité financière et les flux de capitaux. Les données et la méthodologie du travail sont avancées lors de la deuxième section. Nous analysons, interprétons les résultats empiriques et nous discutons leurs implications dans les deux dernières sections.

## 2 | CADRE D'ANALYSE DE L'INTERACTION ENTRE LA STABILITÉ FINANCIÈRE ET LES FLUX DE CAPITAUX

L'interaction entre les flux de capitaux et la stabilité financière est l'élément fondamental de ce travail.

### 2|1 ANALYSE DE FLUX DE CAPITAUX[3]

Les flux des capitaux internationaux sont les transactions financières (cessions et acquisition des actifs et passifs financiers)[4] effectuées entre résidents et non-résidents d'un pays.

### 2|1|1 Composition des flux de capitaux

Pour la composition des flux de capitaux, cette analyse écarte les différentes aides internationales ainsi que les prêts du FMI. Nous analysons les flux privés à savoir les investissements directs étrangers (IDE), les Investissements de portefeuille et les prêts bancaires (ainsi que les dettes à court terme).

---

[3] Il s'agit des flux capitaux financier. Le capital financier est l'épargne disponible pour l'investissement
[4] Les actifs sont des avoirs et les passifs des engagements.

## 2|1|1|1 Les investissements directs étrangers

Généralement, les flux d'IDE ne posent pas trop des problèmes aux économies, puisqu'ils sont moins volatiles.

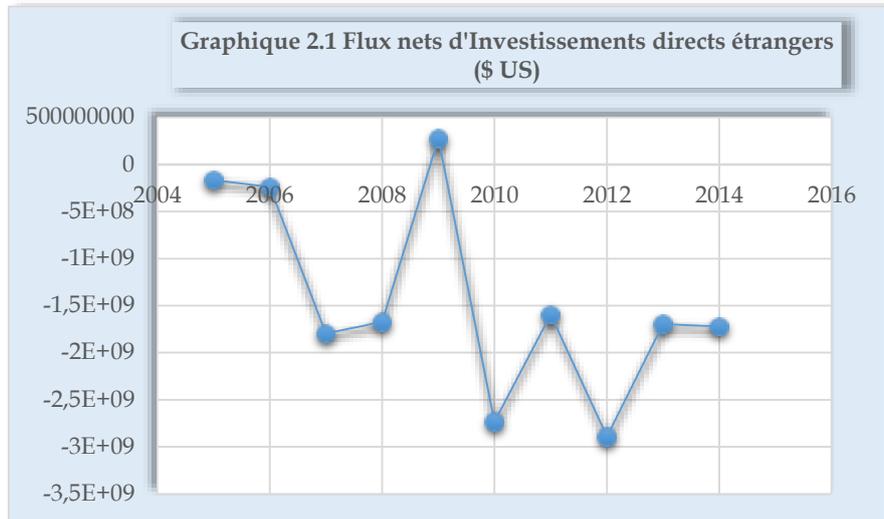

*Source : Banque mondiale, base de données (2015)*

Depuis 2005, les flux nets d'investissements directs étrangers en RDC connaissent une baisse (graphique 2.1). En moyenne de 2005 à 2008 les IDE nets se sont détériorés affichant un solde de 967,9 millions de dollars. Ces flux nets ont montré une baisse drastique juste après la crise financière, en moyenne de 2010 à 2014 les flux net d'investissements directs étrangers ont baissé de 2,1 milliards de dollars.

## 2|1|1|2 Les investissements de portefeuille

Les investissements de portefeuille vers la RDC ont progressé à partir de 2005 passant par trois périodes de baisse.
Le déclenchement de la première période de baisse (2007-2009) pourrait être causé par la forte dépréciation de la monnaie et la crise financière internationale. La deuxième période de baisse (2010-2011), les perturbations politiques ont causé cette baisse d'afflux des investissements de portefeuille. La troisième période (2012-2013) a été marquée par une chute spectaculaire des entrées de capitaux (graphique 2.2). De 2005 à 2014, les flux nets d'investissements de portefeuille[5] ont connu une augmentation de 1,8 milliard de dollars en 2007, en 2010 une augmentation de 3,2 milliards puis en 2012 un niveau record de 3,5 milliards. Il faut retenir que ces flux nets n'ont pas des valeurs négatives, ce qui veut dire que de 2005 à 2014 les flux entrants ont été toujours supérieurs aux flux sortants.

---

[5] L'afflux des investissements de portefeuille peut fournir aux pays en développement une source de financement étranger en substitution de l'endettement extérieur. En effet, ces pays étant en besoin de financement, les investissements de portefeuille viennent compléter l'épargne intérieure destinée au financement de l'investissement.

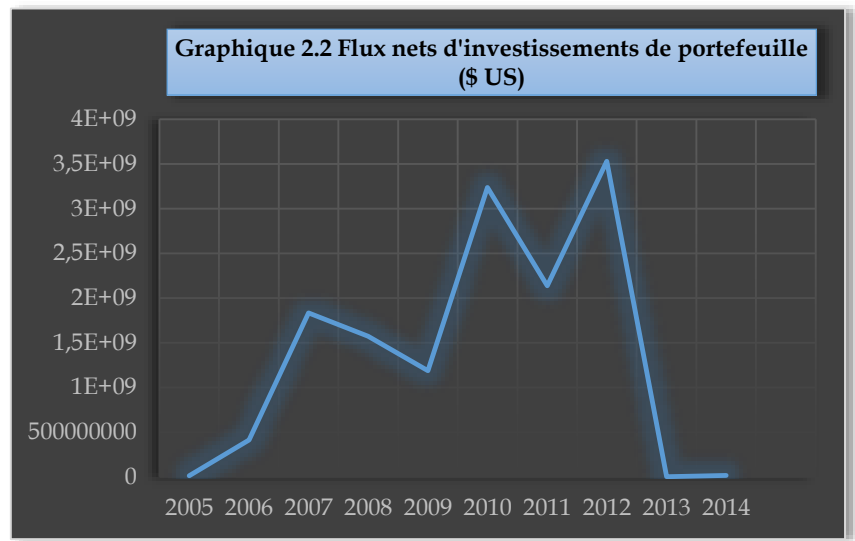

*Source : Banque mondiale, base de données (2015)*

### 2|1|1|3 Les prêts bancaires

Les prêts bancaires, ce sont le financement extérieur assuré par les banques.

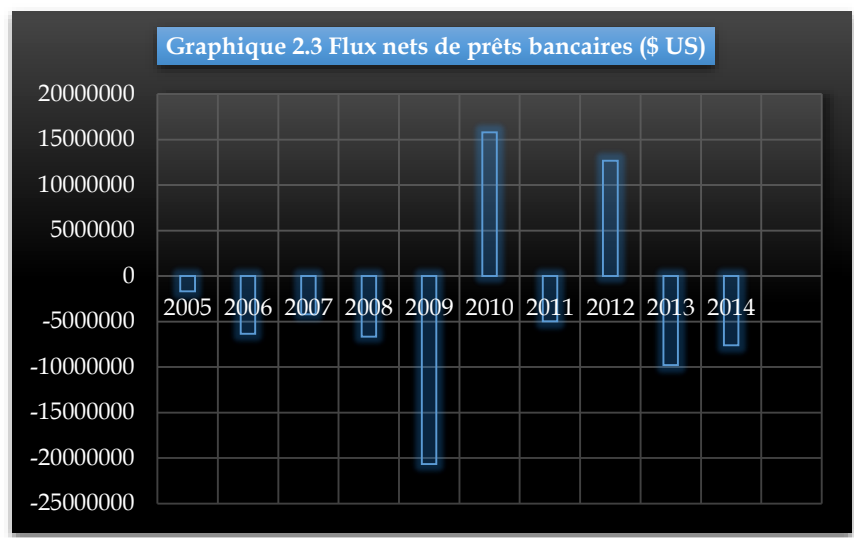

*Source : Banque mondiale, base de données (2015)*

De 2005 à 2009, le pays a connu une sorties de capitaux, avec un niveau de 20,6 millions de dollars, puis il y a eu une volatilité des flux de capitaux de 2010 à 2014 (graphique 2.3). Les prêts bancaires à court terme ont été à l'origine des déséquilibres macrofinanciers qui ont joué un rôle important dans les crises financières sur les marchés émergents dans les années 90 (FMI, 2013).

Les responsables des politiques publiques s'inquiètent plus de la forme exacte prise par les flux de capitaux. On considère en général que les IDE ont des propriétés qui les rendent préférables à la dette. Ils tendent à être moins volatils et à engendrer des avantages indirects comme les transferts de technologie.

D'une manière générale, les IDE et les investissements en capitaux propres posent un peu moins de problèmes que la dette. Mais il ne faudrait pas attacher trop d'importance à cette question. Puisqu'en pratique, les trois types de flux de capitaux sont souvent liés entre eux (ainsi, des firmes étrangères introduiront souvent la trésorerie dans un pays avant d'y acheter effectivement une usine) (Reinhart et Rogoff, 2010).

Les flux de dette s'avèrent être très instables (graphique 2.4) est peut menacer la stabilité du système financier. Ces flux se comportent de manière volatile.

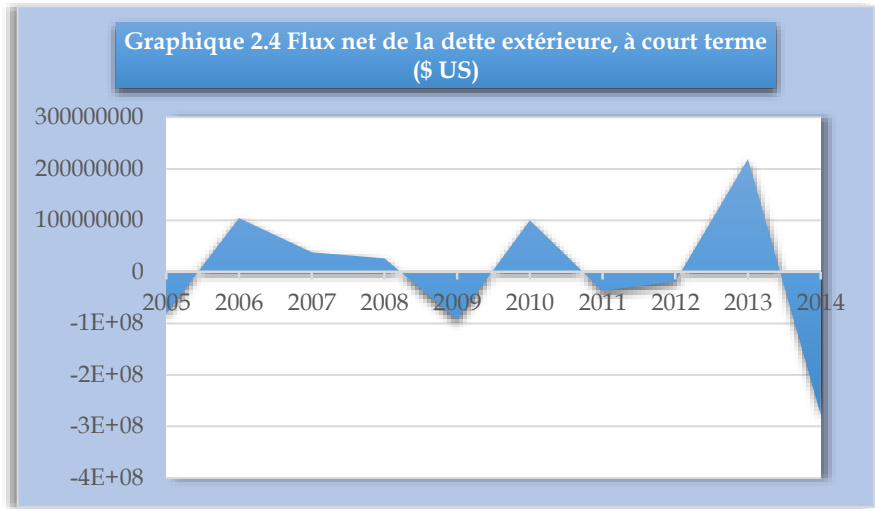

*Source : Banque mondiale, base de données (2015)*

Cependant, Reinhart et Rogoff tirent l'attention en disant que la dette à court terme, classiquement considérée comme la plus susceptible de précipiter des crises de la dette, facilite le commerce international et est nécessaire dans une certaine mesure pour que les agents privés puissent exécuter des stratégies de couverture.

## 2|2 ANALYSE DE LA STABILITÉ FINANCIÈRE

La stabilité financière est devenue à ces jours l'un des objectifs principaux de la politique macroéconomique à la suite de conséquences dévastatrices des épisodes d'instabilité financière. La stabilité financière est une situation qui correspond à un bon fonctionnement ou la robustesse du système financier (Marchés, institutions et infrastructures).

### 2|2|1 Composante du système financier

Le système financier est l'ensemble des institutions financières, des marchés financiers et des infrastructures financières qui dirigent les capitaux des agents à capacité de financement (épargne) vers ceux à besoins de financement (investissement).

- **Les institutions financières**

Les institutions financières tiennent un rôle prépondérant dans tout système financier. Ici les pourvoyeurs de capitaux et les demandeurs de capitaux sont reliés via les institutions financières. Ces institutions regroupent les banques, les compagnies d'assurance, les microfinances, etc. Ce mode de financement est appelé, finance indirecte ou intermédiée. En RD. Congo, les banques commerciales sont les institutions financières dominantes en termes de taille financière (FMI, 2014).

- **Les marchés financiers**

Les marchés financiers sont les lieux sur lequel s'échangent les capitaux. Les marchés financiers se subdivisent en deux marchés de capitaux : le marché de capitaux à long terme (maturité à plus d'un an) appelé marché financier au sens strict et le marché de capitaux à court terme (échéance en moins d'un an) appelé marché monétaire.
Le marché financier au sens strict se compose généralement du marché des actions (ou marché boursier) et du marché obligataire. Le marché monétaire se subdivise en marché interbancaire et en marché des titres de créances négociables. En RDC il y a un marché monétaire subdivisé en marché en banque où les banques se refinancent auprès de la Banque centrale et en marché interbancaire où les banques se prêtent entre elles.

- **Les infrastructures financières**

Par infrastructures financières, on entend l'ensemble des dispositifs réglementaires et techniques qui régissent les flux des capitaux et facilitent leurs utilisations à divers fin. Les principales sont les infrastructures légales, les systèmes d'informations et comptables, les systèmes de paiements, de règlements et de compensation, ainsi que les dispositifs de gestion des crises. L'infrastructure financière c'est la pièce maitresse de la stabilité du système financier. Elle facilite avec sécurité les échanges et maintient la stabilité financière grâce à une régulation saine et solide.

C'est à travers l'infrastructure financière que les autorités atténuent la volatilité des flux de capitaux et supervisent de manière efficace le système financier. Une mauvaise gestion de l'infrastructure financière perturbe la stabilité financière et dégrade l'économie. Pour rendre le système financier plus prudent face aux risques, la banque de règlements internationaux (BRI) recommande de règles prudentielles appelées les accords de Bâle.

## 2|2|2 Profil du système financier congolais

Les pays sur le plan financier présentent une grande variété de situation, allant des pays Développés financièrement (où les systèmes financiers sont très profonds) aux pays en démarrage financier (où les systèmes financiers sont peu profonds et sous-développés).

Le tableau 2.1 dresse le critère financier du plus profond au moins profond. Le manque de profondeur caractérise le profil du système financier congolais (graphique 2.5). En analysant les indicateurs de profondeur du système financier en RD. Congo tels que, le ratio crédit bancaire au secteur privé/PIB et le ratio passif liquide[6]/PIB, on remarque qu'ils demeurent à un niveau très bas. Le premier ne dépasse pas 1% du PIB et le second atteint seulement 6,2% du PIB en 2014 (graphique 2.5). En comparant par rapport au critère, le système financier congolais est en démarrage financier.

---

[6] Les passifs liquides désignent le total de dépôts bancaire (dépôts à vue + dépôts à terme, en CDF et USD).

| Tableau 2.1 | Indicateurs de profondeur du système financier[7] |
|---|---|
| **Credit bancaire au secteur privé/PIB (en %)** | |
| Pays développés | Supérieur à 70 |
| pays émergents | Au moins 60 |
| pays en décollage | Au moins 20 |
| pays en démarrage | Inférieur à 20 |
| **Passifs liquides/PIB (en %)** | |
| Pays développés | Supérieur à 55 |
| pays émergents | Au moins 45 |
| pays en décollage | Au moins 25 |
| pays en démarrage | Inférieur à 25 |
| **Credit bancaire au secteur privé/Dépôt (en %)** | |
| Pays développés | Supérieur à 95 |
| pays émergents | Au moins 80 |
| pays en décollage | Au moins 60 |
| pays en démarrage | Inférieur à 60 |

*Source : Calcul de l'auteur*

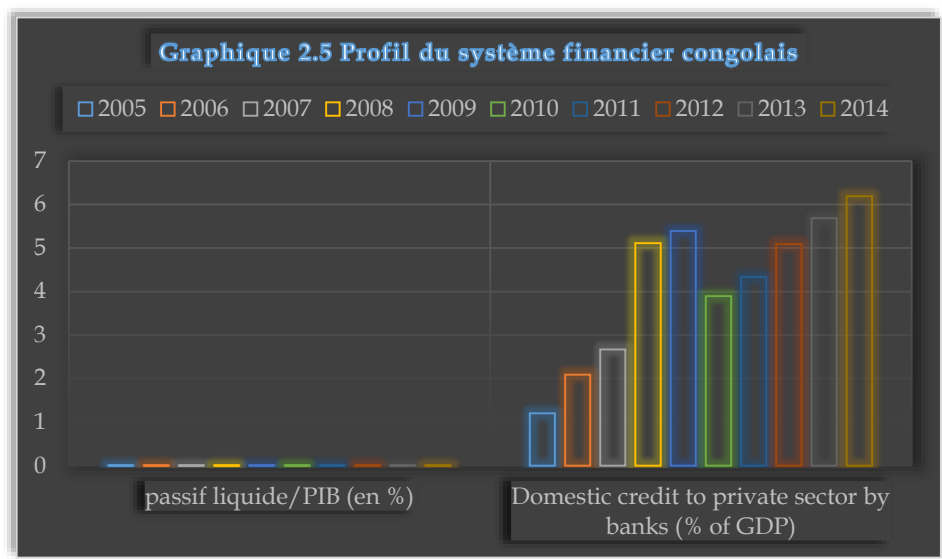

*Source : BCC et Banque Mondiale, calcul de l'auteur*

---

[7] Calcul fait en moyenne sur base d'information de Mohanty (2014), « Le rôle des banques centrales dans la stabilité macroéconomique et financière »

Ce profil montre que le système financier congolais est encore très jeune et peu profond. Dans un contexte des flux massifs de capitaux et de libre circulation de ces derniers, ce manque de profondeur financier pourrait générer une volatilité accrue de capitaux et perturber la stabilité du système financier.

## 2|3 VOLATILITÉ DES FLUX DE CAPITAUX ET STABILITÉ FINANCIÈRE

Cette section est consacrée à l'analyse de la boucle rétroactive ou de l'interaction de la volatilité des flux de capitaux et la stabilité financière.

### 2|3|1 Canaux d'interaction

Le graphique 2.6, un diagramme des flux de capitaux, montre les interactions entre les flux de capitaux et le système financier. Nous remarquons que le système financier dirige les flux financiers de ceux qui ont un excédent de capitaux vers ceux qui ont un besoin.

**Graphique 2.6 Diagramme des flux de capitaux**

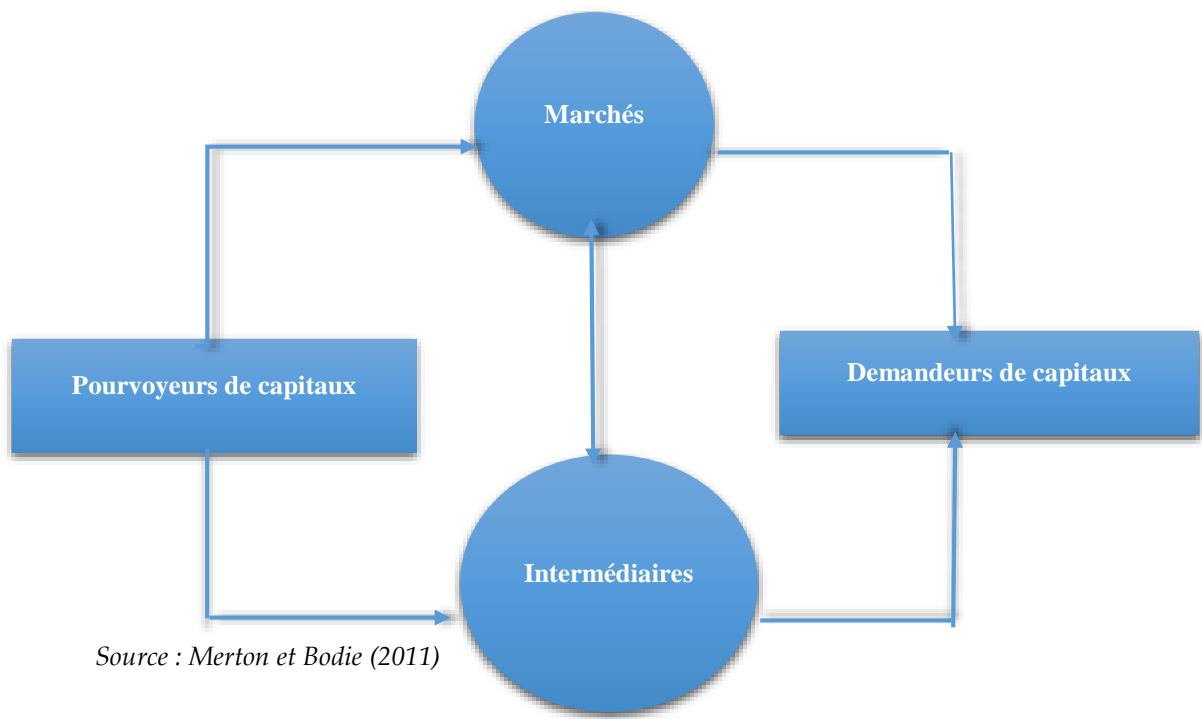

*Source : Merton et Bodie (2011)*

Ce diagramme peut aussi nous aider à montrer comment une volatilité des flux de capitaux peut dysfonctionner la stabilité financière et vice versa. Une instabilité des flux de capitaux perturbera le transit des capitaux c'est-à-dire, le système financier sera en difficulté de transférer les flux financiers des pourvoyeurs vers les demandeurs. Ceci aura comme conséquence la baisse de la croissance économique et l'instabilité financière. De même un système financier moins règlementé, moins solide et peu profond peut perturber les flux de capitaux et les rendre volatiles.

## 2|3|2 Bonanza de capitaux, boom de crédit et sudden stop

L'un des aspects communs des périodes préalables à l'instabilité est un gonflement des entrées de capitaux, c'est ce qu'on appelle un « *bonanza de capitaux* ». La littérature montre qu'une instabilité financière est plus conditionnelle à un bonanza des capitaux. Un bonanza des capitaux est un important indicateur de la vulnérabilité financière. Souvent le bonanza des capitaux est calculé comme un taux de déficit des paiements courants par rapport au PIB. Le seuil de bonanza[8] varie selon les pays. Par exemple, dans les pays relativement fermés, le bonanza est défini comme un ratio déficit de la balance courante/PIB supérieur à 2%, tandis que pour les pays plus ouvert aux échanges internationaux, la limite est un taux de 6,6% (Reinhart et Rogoff, 2010).

Pour le cas de la RD. Congo, nous ne définissons pas la limite, mais la situation des bonanzas assez élevés sont un grand danger pour la stabilité financière du pays. Les bonanzas sont respectivement de 10.6%, 9.1% et 7.6% de 2013 à 2015 (tableau 2.2). Ceci nous montre le degré élevé de la vulnérabilité.

| Tableau2.2 | Bonanza de capitaux |
|---|---|
| **Année** | **Ratio déficit courant /PIB (%)** |
| 2008 | -0,8 |
| 2009 | -6,1 |
| 2010 | -10,5 |
| 2011 | -5,2 |
| 2012 | -6,1 |
| 2013 | -10,6 |
| 2014 | -9,1 |
| 2015 | -7,6 |

*Source : FMI, Perspective de l'économie mondiale (2015)*

Ces constats relatifs aux bonanzas des flux de capitaux sont aussi cohérents avec les « *booms du crédit* » (un phénomène pendant lequel le crédit bancaire au secteur privé s'accroît plus rapidement qu'à son rythme normal).

Le boom de crédit et l'expansion de la liquidité sont fortement corrélés avec les épisodes de bonanzas. La raison est que ces flux (notamment ceux qui prennent la forme de dépôts à court terme en devises) sont souvent intermédiés directement par le système bancaire (Agenor, 2000). Néanmoins, si ces entrées massives des capitaux ont eu lieu dans le cadre d'un système bancaire peu réglementé et moins profond, elle peut engendrer une augmentation des crédits accordés au secteur privé sans pour autant qu'il y ait un niveau d'investissement important ni de projets rentables (Nabila, 2015).

---

[8] On utilise le seuil pour savoir quand est-ce qu'un bonanza peut détériorer le système financier et l'économie.

Le risque d'un bonanza est son renversement brutal qui peut se traduire au « *sudden stop* », c'est-à-dire un arrêt brusque d'entrées de capitaux ou soit un retrait soudain des capitaux ou encore un arrêt brutal des prêts. Ce phénomène est souvent provoqué par l'instabilité macrofinancière et le « *credit crunch*[9] » qui l'accompagne, ou par l'instabilité politique. Le sudden stop à son tour sape la stabilité financière ou exacerbe l'instabilité financière.

# 3 | DONNÉES ET MÉTHODOLOGIE

### 3 | 1 DONNÉES

Les statistiques que nous utilisons couvrent la période mensuelle de 2005 :01 à 2015 :12. Et sont celles publiées par la Banque mondiale, le fonds monétaire international et la Banque centrale. Nous choisissons le ratio credit domestique au secteur privé/PIB, les flux d'IDE, les flux d'investissement de portefeuille, les flux de dette à court terme, le bonanza de capitaux, Le ratio masse monétaire/PIB, le taux d'investissement et le taux d'épargne, le PIB réel par habitant, Le coefficient de réserve obligatoire (CRO) est une variable institutionnelle de la politique monétaire dans l'objectif de réduire l'effet de la volatilité des flux de capitaux sur la stabilité financière (nous utilisons le coefficient des réserves obligatoires de manière discriminatoire, en dollar (CROUSD) et en franc congolais (CROCDF)), Le bon BCC (BBCC) est aussi une variable institutionnelle de la politique monétaire dans le but de stériliser les flux de capitaux, La variable cyclique de l'actif bancaire /réserve (CFPC) est un proxy du coussin de fonds propres contracyclique (Pinshi, 2017) c'est une variable de la politique macroprudentielle et Les variables des flux nets de capitaux (FNCPIB, FNPIB et DNF).

### 3 | 2 MÉTHODOLOGIE

En vertu de l'implication du cadre analytique ci-haut exposé, selon lequel la volatilité des flux de capitaux et la stabilité du système financier interagissent, nous allons mener notre analyse dans le cadre : d'une analyse de la régression linéaire dynamique pour mesurer la taille absolue du système financier, ceci est important pour mesurer le poids systémique du système financier par rapport à l'économie réelle ; d'un modèle de régression dynamique à la Feldstein-Horioka pour tester le degré de mobilité de capitaux ; d'une analyse de la causalité au sens de granger pour vérifier la boucle rétroactive entre la volatilité des flux de capitaux et la stabilité financière et d'une analyse de la décomposition de variance pour mesurer les effets des politiques macrofinancières dans la contribution de la variance prévisionnelle et dans l'atténuation de la volatilité des flux de capitaux.

---

[9] Contraction du crédit

### 3|2|1 Analyse de la corrélation

La matrice de corrélation est très utile pour une mesure descriptive de la robustesse de l'association linéaire entre les flux de capitaux et la stabilité financière. Le coefficient de corrélation est toujours compris entre -1 et +1. Une valeur égale à +1 (-1) indique que les deux variables sont parfaitement liées de façon positive (négative).
Des valeurs proches de zéro indiquent que les deux variables ne sont pas linéairement liées (Anderson et al, 2013). Ainsi le coefficient de corrélation peut être calculé de la manière suivante :

$$r = \frac{\sum XY - n\bar{X}\bar{Y}}{\sqrt{[\sum X^2 - n\bar{X}][\sum Y^2 - n\bar{Y}]}}$$

La significativité des coefficients est importante pour rendre l'analyse fiable et robuste. Ainsi le t student est calculé de la manière suivante :

$$t^* = \frac{|r|}{\sqrt{\frac{(1-r^2)}{n-2}}}$$

### 3|2|2 Modèle dynamique de régression

La plupart des modèles statiques produisent des résidus qui présentent une corrélation sérielle, ce qui amène à douter de la fiabilité des paramètres estimés et de l'analyse évaluée. Afin de résoudre ce problème, les économistes ont souvent recours à des *modèles dynamiques*, qui utilisent des valeurs passées et présentes à la fois pour la variable dépendante (et les variables indépendantes) pour expliquer la valeur courante de la variable dépendante (FMI, 2013).

On peut décider alternativement de considérer dès le départ une forme particulière du modèle dynamique en incluant la variable endogène retardée dans un modèle statique. On a donc simplement :

$$y_t = \gamma_0 + \gamma_1 x_t + \gamma_2 y_{t-1} + \mu_t \tag{3.1}$$

Où $y_{t-1}$ est la variable endogène retardée et $\mu_t$ est le terme de l'erreur.

Dans ce modèle, l'hypothèse qui stipule la non dépendance des variables exogènes avec le terme de l'erreur $\mu_t$ n'est plus satisfaite car la variable $y_{t-1}$ dépend de $\mu_{t-1}$ qui est une variable de perturbation et est aléatoire. Cette situation aura un effet néfaste car la solution serait explosive, pour se rassurer de la fiabilité de l'estimation des paramètres il faudrait ajouter l'hypothèse de la stabilité. La condition de la stabilité est donc :

$$|\gamma_2| < 1$$

- **Test du degré de mobilité de capitaux**

Feldstein et Horioka (1980) ont proposé de mesurer le degré de mobilité des capitaux sur base de la corrélation entre l'épargne et l'investissement. Afin d'égaliser l'investissement à l'épargne nous recourons aux identités comptables suivants :

$$PIB = Y = C + I + G + NX \qquad (3.2)$$

À gauche, on a le produit intérieur brut (Y), c'est-à-dire la valeur de tous les biens et services finaux produits dans l'économie. À droite, on a les sources de la demande globale : consommation privée (C) ; investissement privé (I) ; dépenses publiques (G) et exportations nettes (NX).

$$S_p = Y - T - C \qquad (3.3)$$

L'identité (3.3) indique que l'épargne privée est égale à l'écart entre le PIB et les taxes (T) et la consommation.

$$S_g = T - G \qquad (3.4)$$

L'équation (3.4) indique que l'épargne publique est égale à l'écart entre les recettes publiques et les dépenses publiques.

Pour avoir l'épargne intérieure (nationale) on additionne les deux épargnes :

$$S = S_p + S_g = Y - C - G \qquad (3.5)$$

L'équation (3.2) devient : $S - I = NX \qquad (3.6)$

En réarrangeant (3.6), on obtient l'égalité traditionnelle selon laquelle le taux d'investissement[10] est égal au taux d'épargne moins le compte courant en % du PIB :

$$\frac{I}{Y} = \frac{S}{Y} - \frac{NX}{Y} \qquad (3.7)$$

En économie fermée les exportations nettes sont nulles ($NX = 0$) et l'investissement est égal à l'épargne intérieure. La corrélation entre l'épargne et l'investissement est alors égale à 1.

$$\frac{I}{Y} = \frac{S}{Y} \qquad (3.8)$$

Ce qui signifie qu'une variation de l'épargne nationale induit une variation équivalente de l'investissement. Cependant, en économie ouverte avec une parfaite mobilité des capitaux, il n'y a théoriquement plus de corrélation entre l'épargne intérieure et l'investissement national (Cadoret et al.2009). D'une part, l'épargne intérieure répond aux opportunités d'investissement du système financier

---

[10] Le taux d'investissement c'est le rapport de l'investissement sur le PIB et le taux d'épargne c'est le rapport de l'épargne intérieure sur le PIB.

international et, d'autre part, l'investissement national est financé par l'épargne étrangère.

Dans le cadre de notre étude nous estimons un modèle dynamique à la Feldstein-Horioka :

$$I/PIB_t = \beta_0 + \beta_1 S/PIB_t + \beta_2 I/PIB_{t-1} + \varepsilon_t \qquad (3.9)$$

Où $I/PIB_t$ et $S/PIB_t$ représentent le taux d'investissement et d'épargne. $\varepsilon_t$ est une perturbation.

Le paramètre $\beta_1$ mesure le degré de mobilité des capitaux. C'est le coefficient de rétention de l'épargne[11].

Une estimation de $\beta_1$ proche de 0 conduit donc à considérer une forte mobilité des capitaux, dans la mesure où la corrélation entre l'épargne intérieure et l'investissement national est nulle. En revanche, une estimation de $\beta_1$ proche de 1 indique qu'une variation de l'épargne intérieure induit une variation identique de l'investissement national. L'épargne nationale reste à l'intérieure du pays et la mobilité de capitaux est faible.

- **Taille absolue du système financier**

La taille absolue du système financier est mesurée par la contribution du système financier au développement de l'activité économique ou à la croissance économique. C'est-à-dire le pouvoir explicatif ou le degré d'ajustement du système financier à la croissance économique.

Le PIB réel par habitant ($y_t$) sera utilisé comme variable endogène pour mesurer le développement de l'activité économique et/ ou la performance économique. C'est l'indicateur de la sphère réelle. Le ratio crédit domestique accordé au secteur privé /PIB ($Dcp_t$) et la profondeur (ou développement) financier ($M2/y_t$) seront utilisés pour mesurer le système financier.

$$y_t = \gamma_0 + \gamma_1 Dcp_t + \gamma_2 M2/y_t + \gamma_3 y_{t-1} + \varepsilon_t \qquad (3.10)$$

La taille absolue du système financier sera mesurée par le degré d'ajustement ou le pouvoir explicatif du système financier sur l'économie réelle :

$$R^2 = \frac{\sum(\hat{Y}_t - \bar{Y})^2}{\sum(Y_t - \bar{Y})^2} = 1 - \frac{\sum \varepsilon^2}{\sum(Y_t - \bar{Y})^2} \qquad (3.11)$$

### 3|2|3 Test de causalité à la Granger.

L'objectif de ce document du travail est d'analyser les interactions entre la volatilité des flux de capitaux et la stabilité financière, plus précisément il s'agit de vérifier la boucle rétroactive (feedback effect) c'est-à-dire la causalité bidirectionnelle, entre la volatilité des flux de capitaux et la stabilité financière.

---

[11] C'est-à-dire la fraction de la variation exogène d'épargne restant dans le pays et finançant l'investissement national.

Si la volatilité des flux de capitaux cause la stabilité financière, il s'agirait d'un risque pour le secteur financier. En effet si ces flux sont accompagnés des arrêts soudains, et d'autres menaces, ils pourraient se traduire par une perturbation de la stabilité financière et alimenteraient une crise financière.

S'il y a une boucle rétroactive, le système financier sera qualifié du processus de rattrapage financier et d'une voie vers un renforcement de la régulation financière pour atténuer la volatilité des flux de capitaux.

La méthodologie développée par Granger nous permet d'évaluer si la volatilité des flux de capitaux et la stabilité financière ont une causalité bidirectionnelle.

Nous distinguons deux variables, la volatilité des flux de capitaux ($Fc$) et la stabilité financière ($Sf$). Le test de causalité au sens de granger (Granger, 1969) suppose que $Fc$ cause $Sf$ si la prévision de $Sf$ fondée sur l'information passée de $Fc$ et de $Sf$ est meilleure que la prévision fondée uniquement sur l'information passée de $Sf$.

$$Fc_t = \tau_0 + \sum_{i=1}^{p} \tau_i\, Fc_{t-i} + \sum_{i=1}^{p} \xi_i\, Sf_{t-i} + \upsilon_{1t} \qquad t = 1,\ldots,T \qquad (3.12)$$

$$Sf_t = \eta_0 + \sum_{i=1}^{p} \eta_i\, Sf_{t-i} + \sum_{i=1}^{p} \varphi_i\, Fc_{t-i} + \upsilon_{2t} \qquad (3.13)$$

L'hypothèse nulle que $Fc$ ne cause pas $Sf$ consiste à tester la nullité jointe des paramètres :
$H_0 : \xi_1 = \ldots = \xi_p = 0$

L'hypothèse nulle que $Sf$ ne cause pas $Fc$ consiste à tester la nullité jointe des paramètres :
$H_0 : \varphi_1 = \ldots = \varphi_p = 0$

### 3|2|4 Décomposition de la variance

L'objectif secondaire de ce travail est de proposer des politiques pour le renforcement de la gestion de la stabilité financière en analysant la manière à laquelle les politiques macroéconomiques et prudentielles visent à atténuer la volatilité des flux de capitaux. Pour mieux comprendre les contributions des politiques dans l'atténuation de la volatilité des flux de capitaux nous exploitons les propriétés de l'erreur de prévision, notamment de l'information que donne la variance de cette erreur. En effet, la décomposition de la variance de l'erreur de prévision nous sert d'approximation pour quantifier la variabilité des flux de capitaux et par conséquent la volatilité.

Soit un processus VAR ($p$):

$$y_t = m + A_1 y_{t-1} + A_2 y_{t-2} + \cdots + A_p y_{t-p} + \varepsilon_t \qquad (3.14)$$

Où les $A_i$ sont des matrices de coefficients d'ordre $k$, $m$ est un vecteur de $k$ constantes, $\varepsilon_t$ étant un vecteur de bruits blancs, qui a les propriétés suivantes :

$$E(\varepsilon_t) = 0 \text{ pour tout } t \quad E(\varepsilon_t \varepsilon_t') = \begin{cases} \Omega & s = t \\ 0 & s \neq t \end{cases}$$

Supposons que la matrice de variance-covariance $\Omega$ est définie positive. Les $\varepsilon$ ne sont pas autocorrélées mais peuvent être corrélés à un moment donné (Johnston et Dinardo, 1999).

Pour saisir les effets des politiques macroéconomiques et prudentielles dans l'atténuation de la volatilité des flux de capitaux en vue d'atteindre la stabilité financière, il est préférable de reformuler le processus VAR ($p$) au processus VMA ($\infty$). La présentation VMA ($\infty$) de l'équation (3.14) est donnée par :

$$y_t = \mu + v_t + B_1 v_{t-1} + B_2 v_{t-2} + \cdots = \mu + \sum_{i=0}^{\infty} B_i v_{t-i} \qquad (3.15)$$

Avec : $\mu = (I - A_1 - A_2 - \cdots - A_p)^{-1} \times m$

Et $B_i = \sum_{j=i}^{\min(p,i)} A_j B_{i-j} \qquad i = 1, 2, \ldots \text{ et } B_0 = 0$

Sous cette forme, la matrice $B$ apparaît comme un « multiplicateur d'impact », c'est-à-dire que c'est au travers de cette matrice qu'un choc se répercute tout le long du processus. Une modification en un instant donné t de $v_t$ affecte $y_t$ (Bourbonnais, 2015). Soit un processus VAR (1) à deux variables $y_t$ et $x_t$, la variance de l'erreur de prévision pour $y_{t+h}$ s'écrit :

$$\sigma_{y_t}^2(h) = \sigma_{\varepsilon_1}^2 [b_{11}^2(0) + b_{11}^2(1) + \cdots + b_{11}^2(h-1)] + \sigma_{\varepsilon_2}^2 [b_{22}^2(0) + b_{22}^2(1) + \cdots + b_{22}^2(h-1)] \qquad (3.16)$$

Où les $b_{ii}$ sont les termes de la matrice $B$

À l'horizon $h$, la décomposition de la variance ou la contribution (en %) des cadres opérationnels des politiques à la variance prévisionnelle ou la volatilité des flux de capitaux, est donnée par :

$$\frac{\sigma_{\varepsilon 2}^2 [b_{22}^2(0) + b_{22}^2(1) + \cdots + b_{22}^2(h-1)]}{\sigma_{y_t}^2(h)}$$

# 4| RELATION ENTRE LA STABILITÉ FINANCIÈRE ET LA VOLATILITÉ DES FLUX DE CAPITAUX EN RD. Congo

## 4|1 LIEN ENTRE LA STABILITÉ FINANCIÈRE ET LES FLUX DE CAPITAUX

Nous procédons par l'analyse de la corrélation entre les flux de capitaux et les variables mesurant la performance du système financier.

Le ratio credit domestique au secteur privé/PIB (Dcp) mesure le degré d'intermédiation et de pénétration du système financier dans l'économie. C'est l'indicateur par excellence de la stabilité financière[12]. Ce ratio est lié positivement avec les flux d'IDE (IDE) mais d'un faible degré 0,25. Il est faiblement lié de façon négative avec les flux d'investissement de portefeuille (IPOR). Les flux de dette à court terme (DCT) n'a presque pas une relation linéaire avec le crédit bancaire au secteur privé. La relation entre le bonanza de capitaux (Bcap) et le crédit bancaire est modérée (0,38) et positive. Un afflux massif de capitaux peut se traduire par un excès des crédits et déclencher une instabilité financière ou encore un reflux soudain de capitaux pourrait se traduire par une instabilité financière. La vulnérabilité du système financier est bien vérifiée par l'effet bonanza et le sudden stop.

*Tableau 4.1 Matrice de corrélation entre les flux de capitaux et la stabilité financière en RD. Congo*

|        | Dcp     | M2/PIB  | IDE     | IPOR    | DCT     | Bcap    |
|--------|---------|---------|---------|---------|---------|---------|
| **Dcp**    | 1,00    | 0,29    | **0,25\*\*** | -0,14   | **0,01** | **0,38\*\*** |
| **M2/PIB** | 0,29    | 1,00    | **0,18\*\*** | **-0,31\*\*** | **0,62\*\*** | **0,46\*\*** |
| **IDE**    | **0,25\*\*** | **0,18\*\*** | 1,00    | -0,83   | -0,20   | 0,23    |
| **IPOR**   | -0,14   | **-0,31\*\*** | -0,83   | 1,00    | -0,20   | 0,01    |
| **DCT**    | 0,01    | **0,62\*\*** | -0,20   | -0,20   | 1,00    | -0,56   |
| **Bcap**   | **0,38\*\*** | **0,46\*\*** | 0,23    | 0,01    | -0,56   | 1,00    |

*Source : calcul de l'auteur*

Le ratio masse monétaire/PIB (M2/PIB) est souvent utilisé pour mesurer le niveau du développement ou la taille du système financier. Ce ratio est fortement lié (0,62) de façon positive avec le flux de dette à court terme. Les flux de dette à court terme sont les flux le plus volatiles (graphique 2.3 et 2.4). Cette forte relation met en évidence le renforcement de la régulation, puisqu'une instabilité accentuée de ces flux pourrait saper la stabilité financière. De même la liquidité (M2/PIB) a une importante liaison avec le bonanza de capitaux.

L'analyse de la corrélation est une analyse présomptive, la causalité pourrait bien expliquer de façon approfondie ces relations.

## 4|2 TEST DU DEGRÉ DE MOBILITÉ DES CAPITAUX

Il existe une forte corrélation négative entre mobilité des capitaux et stabilité financière. Puisque souvent les périodes de forte mobilité des capitaux ont de manière répétée causé des crises financières (Reinhart et Rogoff, 2010). Certes, une libre circulation des capitaux est souvent liée à un environnement déréglementé. Les banques évoluent dès lors dans un niveau de règlementation et de supervision insuffisant.

Les afflux de capitaux internationaux suite à la globalisation financière[13] sont généralement accompagnés d'un excès d'optimisme quant aux bénéfices futurs.

---

[12] Son avantage est qu'il exclut le crédit au secteur public, met l'accent sur le rôle des institutions financières dans le financement des secteurs productifs.

[13] La globalisation financière désigne une expansion du système financier international induite par la déréglementation financière (abolition ou diminution des restrictions financières).

L'abolition des restrictions à laquelle vient s'ajouter une perception optimiste de la conjoncture économique, conduit souvent à une augmentation de la demande de crédit. Afin de satisfaire une telle demande qui augmente plus rapidement que les dépôts domestiques, les banques peuvent recourir aux capitaux étrangers. Ces capitaux, étant intermédiés directement ou indirectement par les banques domestiques, viendront augmenter la capacité d'octroi du crédit de ces dernières. Les banques se trouvent exposées au risque de *maturity mismatch*[14] (asymétrie des échéances ou décalage de maturité). Par ailleurs, étant de plus en plus dépendante des financements étrangers, les banques courent aussi un risque de *currency mismatch* (asymétrie de devises ou décalage monétaire), c'est-à-dire, le fait que la banque ait un passif en devise supérieur à son actif en devise[15].

En cas de choc subi par l'économie du pays hôte, il peut y avoir un retrait soudain des capitaux étrangers et un arrêt d'octroi de crédit. Un tel phénomène va engendrer un mouvement de panique au niveau des déposants qui vont procéder à un retrait massif et soudain de leurs avoirs en banques.

Cela peut ainsi engendrer la faillite des banques insolvables et/ou illiquides lors d'une telle ruée aux guichets (Nabila, 2015). Cependant, si le système financier du pays est profond et solide, un degré d'ouverture élevé peut ne pas provoquer une instabilité financière. Toutefois, on note que des périodes de forte mobilité des capitaux ont souvent produit des crises. Ceci amène à prôner pour une règlementation saine du système financier.

La mobilité des capitaux est fortement liée à la perturbation de la stabilité financière. Ce test nous aide non seulement à voir le degré de la mobilité des capitaux mais aussi à mesurer l'effet de l'épargne intérieure et extérieure sur l'investissement national. C'est-à-dire à mesurer le lien entre l'épargne et l'investissement.

Cette modélisation se fera à la Feldstein-Horioka. Ces auteurs ont cherché à évaluer le degré de mobilité des capitaux sur la base d'analyse de lien entre le taux d'investissement et le taux d'épargne.

**Tableau 4.2 Test du degré de mobilité des capitaux[16]**

$$I/PIB_t = 0{,}33\ S/PIB_t + 0{,}40\ I/PIB_{t-1}$$
$$(0{,}00) \qquad (0{,}00)$$

$R^2$ ajusté = 69 %
n = 120 observations
( ) sont les valeurs-p

*Source : calcul de l'auteur*

Le coefficient de rétention est 0,33. Ceci conclut que 33% de l'investissement national est financé par l'épargne intérieure alors que 67% par l'épargne étrangère. Ceci implique que l'économie congolaise est mobile.

---

[14] Le décalage de maturité c'est une situation où les dépôts bancaires sont à court terme alors que les prêts sont à long terme.
[15] Les dépôts peuvent être en devise alors que les prêts sont souvent libellés en monnaie nationale
[16] La constante a paru non significative dans l'estimation, d'où son exclusion.

Ce résultat nous montre que l'économie congolaise est largement financée par l'épargne étrangère (c'est-à-dire les flux de capitaux internationaux) via le système financier. Il ne faut pas prendre à la légère la nature volatile des flux de capitaux et l'analyse de la stabilité financière, car un reflux massif ou un sudden stop peut facilement créer une crise financière. D'où la nécessité d'une régulation renforcée.

**4|3 ANALYSE DE LA TAILLE DU SYSTÈME FINANCIER**

La taille du système financier est la grandeur ou l'importance du système financier.
La grandeur du système financier se mesure du point de vue relative et absolue (Eggoh, 2009).

### 4|3|1 Taille relative du système financier

La taille relative du système financier est mesurée par les différents entités et ratios du système financier.
Dans la deuxième section, nous avons vu que le système financier congolais n'est pas profond, c'est-à-dire sa taille relative est petite (cf. Tableau 2.1 et graphique 2.5).
La taille relative peut être aussi mesurée par le service financier. Nous remarquons que le service financier est à son début expansif (graphique 4.1), les nombres de succursales bancaires et les prêts bancaires augmentent à un rythme lent. Mais le constat est que les épargnants de 2005 à 2014 ont augmenté spectaculairement, bien que la taille soit encore faible.

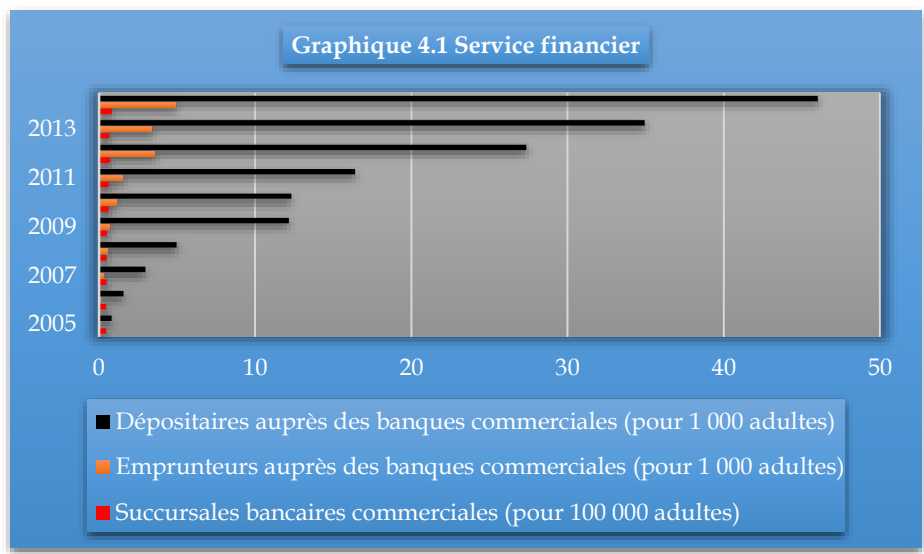

*Source : Banque mondiale (2015)*

### 4|3|2 Taille absolue du système financier

La taille absolue du système financier est mesurée par la contribution du système financier au développement de l'activité économique. C'est-à-dire le pouvoir explicatif ou le degré d'ajustement du système financier à la croissance économique.

Le système financier contribue au développement économique et à l'activité à 46%. Le degré d'ajustement est moyen, ceci veut simplement dire que le système financier a un pouvoir moyen sur l'économie réelle.

Contrairement à la taille relative, l'évaluation de la taille absolue a montré un poids systémique du système financier vis-à-vis de l'économie réelle. Ceci implique qu'un choc du système financier peut devenir systémique et causer des effets dévastateurs à la sphère réelle.

**Tableau 4.3 Estimation de la taille absolue du système financier**

| variables indépendantes et constante | Paramètres estimés | Écart-type | T de student |
|---|---|---|---|
| crédit au secteur privé | -0,25*** | 0,08 | -3,14 |
| Profondeur financier | 0,25*** | 0,49 | 5,07 |
| PIB per capita retardé | 0,36*** | 0,08 | 4,14 |
| constante | -0,63*** | 0,22 | -2,77 |

R² ajusté = 46%

*Source : calcul de l'auteur*

## 4|4 BOUCLE RÉTROACTIVE ENTRE VOLATILITÉ DES FLUX DE CAPITAUX ET LA STABILITÉ FINANCIÈRE

Dans toutes les sections précédentes, nous avons mis en lumière le lien entre la volatilité des flux de capitaux et la stabilité financière. Cependant, l'évaluation qualitative et quantitative de l'interaction entre la volatilité des flux de capitaux et la stabilité financière se fera par l'analyse de causalité.

En effet s'il y a effet feedback entre la volatilité des flux de capitaux et la stabilité financière, il s'agirait d'un phénomène de rattrapage financier et d'une voie vers un renforcement de la régulation financière pour atténuer la volatilité des flux de capitaux. Au contraire, si la volatilité des flux de capitaux cause la stabilité financière, alors ce serait le signe d'un risque potentiel pour le système financier où les flux de capitaux seraient accompagnés des arrêts soudains, et d'autres menaces, qui pourraient se traduire par une perturbation de la stabilité financière et alimenteraient une crise financière.

Le résultat est résumé dans le graphique suivant (les résultats détaillés sont présentés en annexe).

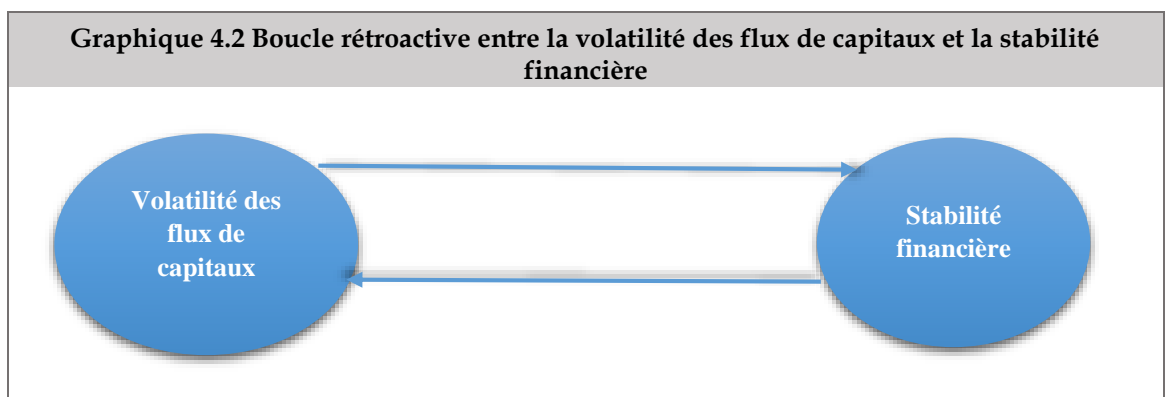

Graphique 4.2 Boucle rétroactive entre la volatilité des flux de capitaux et la stabilité financière

*Source : Calcul de l'auteur*

Le résultat du test montre qu'il existe une boucle rétroactive entre la volatilité des flux de capitaux et la stabilité financière, c'est –à-dire qu'il y a interaction. Ce qui, à la lumière du test de causalité, peut être interprété comme une voie pour renforcer la régulation des politiques visant à atténuer la volatilité des flux de capitaux et traduire un processus de rattrapage et/ou de développement financier.

# 5| POLITIQUES D'ATTÉNUATION DES EFFETS DE LA VOLATILITÉ DES FLUX DE CAPITAUX SUR LA STABILITÉ FINANCIÈRE

La section précédente nous a permis de vérifier l'existence d'une boucle rétroactive entre la volatilité des flux de capitaux et la stabilité financière. Cette évidence donne accès à une voie pour les politiques macroéconomique et macroprudentielle susceptibles d'atténuer les effets de la volatilité de capitaux sur la stabilité financière.

## 5|1 CONDITIONS PRÉALABLES AUX GAINS DES FLUX DE CAPITAUX

L'analyse du degré de mobilité des capitaux à la Feldstein-Horioka nous a montré que l'économie congolaise est mobile financièrement, cela implique que, les flux de capitaux circulent avec moins de restrictions, traduisant par ailleurs une dépendance du système économique et financier à l'épargne étrangère

**Graphique 5.1 Conditions préalables aux gains des flux de capitaux**

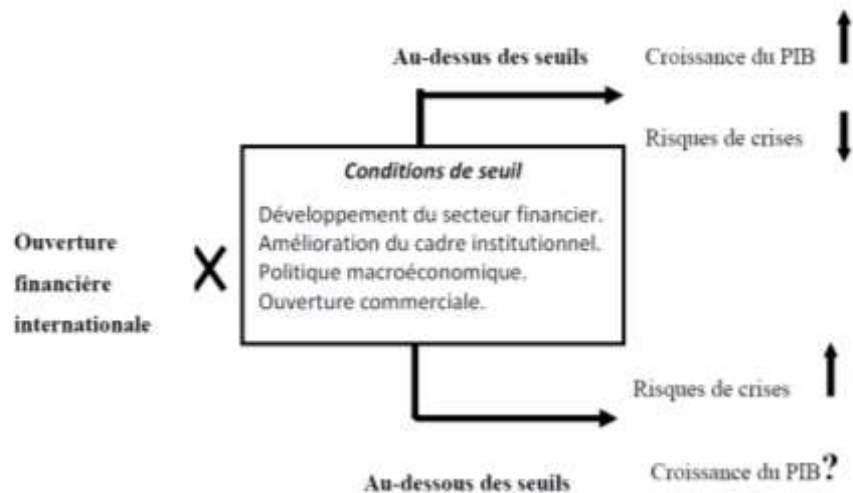

*Source : Kose et al. (2007)*

Les flux de capitaux s'accompagnent de toute une série d'avantages susceptibles d'améliorer l'économie. Toutefois, les pays en développement doivent remplir certaines conditions essentielles pour se protéger contre le sudden stop et les crises financières. Pour réduire la vulnérabilité et bénéficier des gains des flux de capitaux, il conviendrait de réunir quelques conditions minimales dans un pays.

Le graphique 5.1 nous montre que si les flux de capitaux sont conditionnés par un développement du système financier (y compris le renforcement de la régulation), une amélioration du cadre institutionnel (y compris la bonne gouvernance), une bonne politique macroéconomique et une ouverture commerciale, ces flux se traduiraient par une croissance économique durable et par une stabilité financière. Dans ce cadre, les pays seraient au-dessus du seuil. Cependant, si les flux de capitaux ne sont pas conditionnés par ces fondamentaux, c'est-à-dire au-dessous du seuil, cela pourrait se traduire par une crise.

## 5|2 POLITIQUES MACROÉCONOMIQUE ET MACROPRUDENTIELLE

La volatilité des flux des capitaux est un phénomène inévitable, mais dont l'ampleur peut être atténuée. Les politiques macroéconomiques et macroprudentielles peuvent atténuer ses effets perturbateurs. Dans la section 5|4 nous allons évaluer par l'analyse de la décomposition de la variance, les effets de politiques macroéconomiques et macroprudentielles sur la volatilité des flux de capitaux.

### 5|2|1 Politique macroéconomique

La politique macroéconomique est généralement subdivisée en politique monétaire et budgétaire. Une intervention de politique macroéconomique peut atténuer les effets de la volatilité. Les observations empiriques ont montré que la politique monétaire semble bien efficace dans la gestion des flux de capitaux, mais rien ne permet de conclure à l'existence d'un lien systématique entre les flux de capitaux et l'efficacité de la politique budgétaire (Kose et al. 2007).

Dans le cadre de notre analyse, nous ferons plus attention à la politique monétaire, en raison de son influence sur la stabilité financière. Cependant, il sied de souligner qu'en l'absence d'une politique budgétaire solide et disciplinée, il est difficile pour la politique monétaire d'être efficace.

La politique monétaire peut, via son cadre opérationnel[17], atténuer les effets de la volatilité des flux de capitaux, par une intervention stérilisée[18]. Cette stérilisation est efficace si la mobilité de capitaux est moins que parfaite (Agenor, 2000). Il y a d'autres instruments tels que le taux directeur et les réserves obligatoires qui peuvent alléger les effets sur la liquidité et les tensions inflationnistes inhérentes, impulsées par l'expansion des afflux de capitaux.

### 5|2|2 Politique macroprudentielle

Les insuffisances des politiques microprudentielle et macroéconomique à assurer la stabilité financière ont conduit à renforcer les outils de la régulation financière. L'orientation macroprudentielle analyse le système financier dans son ensemble en limitant l'effet du risque systémique. De Bandt et al. (2013) définissent le risque systémique comme un événement systémique défavorable affectant une large part du système financier, conduisant à une dégradation partielle ou totale, mais toujours significative, de sa capacité à assurer ses fonctions fondamentales, à savoir d'offrir des services financiers, avec des conséquences macroéconomiques ou des effets sur le bien-être.

---

[17] Ensembles d'instruments de la politique monétaire visant à réguler l'économie.
[18] L'intervention stérilisée consiste à échanger directement des titres contre les entrées et sorties de capitaux.

La politique macroprudentielle peut être utilisée pour limiter l'ampleur de la volatilité des flux de capitaux (Bernanke, 2011 ; Caruana et Cohen, 2014). Selon Kim (2014), l'élaboration d'un cadre macroprudentiel, en particulier, devrait limiter la procyclicité des flux de capitaux et rendre le système financier contracyclique via son instrument de fond propre contracyclique. Ce coussin de fond propre contracyclique a pour objet de constituer ou d'augmenter les fonds propres des institutions financières en période d'expansion pour se doter d'un volant contracyclique en période de récession (caractérisée par un credit crunch qui provoque un sudden stop).

## 5|3 CONTROLE DES CAPITAUX

Les flux de capitaux constituent l'un des éléments importants du déclenchement des crises financières. La plupart des pays émergents et en développement ont vécu plusieurs épisodes de boom-bust ou d'afflux massif et de retrait soudain de ces capitaux. Le contrôle des capitaux est une ligne défensive des politiques pour limiter la mobilité des capitaux et éviter l'instabilité financière.

Hormis les cadres opérationnels de la politique monétaire et de la politique macroprudentielle, les responsables peuvent utiliser les instruments de contrôle des capitaux pour atténuer la volatilité des flux de capitaux. Il existe plusieurs approches pour contrôler les flux de capitaux notamment : les taxes et les impôts sur les transactions financières (par exemple la taxe tobin); les contrôles des changes ; etc. Il faut noter que tous les flux de capitaux n'ont pas le même degré de volatilité (les IDE sont moins volatiles et beaucoup plus stable que les autres flux de capitaux).

Les flux de dettes à court terme et quelques autres flux à long terme rendent les pays vulnérables (Rogoff, 2002). D'autres instruments d'atténuation de la volatilité des flux de capitaux sont : l'accumulation de réserves, flexibilité du taux change ; la discipline macroéconomique, etc.

Cependant, les contrôles sur les sorties et les entrées de capitaux peuvent développer la corruption. Il est difficile de se prononcer sur l'efficacité des mesures de contrôle des flux de capitaux dans l'environnement actuel, où le commerce est libéralisé et où des nombreux instruments financiers permettent d'y échapper (Mishkin, 2013).

## 5|4 ANALYSE DE LA DÉCOMPOSITION DE LA VARIANCE

L'analyse de la décomposition de la variance aide à déterminer la contribution (en %) des cadres opérationnels de la politique monétaire et de la politique macroprudentielle à la variance prévisionnelle ou la volatilité des flux de capitaux. Une contribution d'un degré élevé des cadres opérationnels de la politique monétaire et de la politique macroprudentielle impliquerait l'efficacité et l'aptitude de ces politiques dans l'atténuation de la volatilité des flux de capitaux. Au cas contraire, c'est-à-dire, un faible degré de contribution impliquerait l'inefficacité et la non aptitude de ces politiques dans l'atténuation.

**Tableau 5.1 Décomposition de la variance des flux net de capitaux (%PIB)**

| Période | S.E. | FNCPIB | FNPIB | DNF | BBCC | CROUSD | CROCDF | CFPC |
|---|---|---|---|---|---|---|---|---|
| 1 | 0,58 | 100,00 | 0,00 | 0,00 | 0,00 | 0,00 | 0,00 | 0,00 |
| 2 | 0,59 | 99,19 | 0,20 | 0,01 | 0,38 | 0,01 | 0,20 | 0,01 |
| 3 | 0,61 | 97,69 | 0,54 | 0,05 | 0,91 | 0,02 | 0,49 | 0,30 |
| 4 | 0,62 | 96,17 | 0,82 | 0,07 | 1,54 | 0,02 | 0,50 | 0,88 |
| 5 | **0,63** | **94,58** | **1,09** | **0,10** | **2,07** | **0,02** | **0,50** | **1,64** |
| 6 | 0,63 | 93,09 | 1,31 | 0,11 | 2,49 | 0,03 | 0,50 | 2,48 |
| 7 | 0,64 | 91,76 | 1,50 | 0,12 | 2,79 | 0,04 | 0,49 | 3,31 |
| 8 | 0,64 | 90,61 | 1,66 | 0,12 | 2,98 | 0,04 | 0,48 | 4,10 |
| 9 | 0,64 | 89,63 | 1,80 | 0,12 | 3,09 | 0,05 | 0,48 | 4,82 |
| 10 | 0,65 | 88,79 | 1,93 | 0,12 | 3,15 | 0,06 | 0,48 | 5,48 |
| 11 | 0,65 | 88,06 | 2,04 | 0,12 | 3,17 | 0,07 | 0,47 | 6,07 |
| 12 | **0,65** | **87,43** | **2,14** | **0,12** | **3,17** | **0,08** | **0,47** | **6,60** |
| 13 | 0,65 | 86,88 | 2,23 | 0,13 | 3,15 | 0,09 | 0,47 | 7,06 |
| 14 | 0,66 | 86,39 | 2,30 | 0,14 | 3,13 | 0,09 | 0,46 | 7,47 |
| 15 | 0,66 | 85,96 | 2,37 | 0,16 | 3,12 | 0,10 | 0,46 | 7,84 |
| 16 | 0,66 | 85,56 | 2,43 | 0,18 | 3,11 | 0,11 | 0,46 | 8,15 |
| 17 | 0,66 | 85,20 | 2,48 | 0,21 | 3,11 | 0,11 | 0,46 | 8,43 |
| 18 | **0,66** | **84,87** | **2,52** | **0,24** | **3,12** | **0,12** | **0,46** | **8,67** |

*Cholesky Ordering: FNCPIB FNPIB DNF BBCC CROUSD CROCDF CFPC*

*Source : Calcul de l'auteur*

Les résultats nous renseignent qu'au 5ème mois, la variance de l'erreur de prévision ou la volatilité des flux de capitaux est due à 2,07% aux innovations ou aux chocs du bon BCC, à 0,5% aux innovations des réserves obligatoires et à 1,6% aux innovations du coussin des fonds propres contracyclique.

Au 12ème mois, la contribution du coussin des fonds propres est à 6,6%, par contre celle du bon BCC est à 3,1% et celle du coefficient des réserves obligatoires est à 0,4%.

Au 18ème mois, 8,6% de la volatilité des flux est due aux innovations du coussin de fonds propres contracyclique, 3,1% et 0,4% aux innovations du bon BCC et de coefficient des réserves obligatoires.

La contribution de la politique macroprudentielle à la variance prévisionnelle est mieux que celle de la politique monétaire. Un choc ou une innovation de la politique macroprudentielle a un petit impact sur la volatilité des flux de capitaux alors que le choc ou l'innovation de la politique monétaire n'as pas assez d'impact sur la volatilité.

Cette analyse nous approxime à dire que les effets de la politique macroprudentielle pourraient atténuer la volatilité des flux de capitaux mieux que ceux de la politique monétaire. D'où, l'adoption d'une approche macroprudentielle semble être plus appropriée dans la gestion de la volatilité des flux de capitaux.

Cependant, le degré d'atténuation des politiques est faible pour juguler la volatilité des flux de capitaux. Ceci implique un renforcement de la régulation financière et un développement du système financier en vue de renforcer sa capacité à absorber les chocs provoqués par une volatilité des flux de capitaux.

## 5|5 RENFORCEMENT DE LA RÉGULATION DU SYSTÈME FINANCIER

L'évaluation dynamique de la décomposition de la variance a donc montré que les politiques macroéconomiques, à l'heure actuelle, ne peuvent pas suffisamment atténuer la volatilité des flux de capitaux. Toutefois, cette évaluation a quand même fourni une piste essentielle sur l'adoption d'une approche macroprudentielle laquelle semble avoir un petit effet dans l'atténuation de la volatilité des flux de capitaux.

La nécessité de renforcer l'orientation macroprudentielle des dispositifs de régulation et de surveillance demeure une priorité pour la bonne santé financière et réelle (Pinshi, 2017). La Banque Centrale, en tant que régulateur de la stabilité financière, doit axer sa politique financière vers une approche macroprudentielle en se dotant des indicateurs avancés pour être capable de prévoir et de résorber un choc systémique menaçant le système financier et l'économie réelle.
Il conviendrait également développer l'infrastructure financière, notamment avec l'implantation des mécanismes d'assurance dépôt dans le but de renforcer la confiance du public et l'épargne intérieure pour promouvoir un approfondissement financier. Cette action permet une meilleure diversification des portefeuilles et des risques, ce qui améliore en fin de compte la résistance du système financier aux chocs et à la volatilité des flux de capitaux (Nasution, 2011).

Dans une approche prospective et préventive, le régulateur devrait se doter des indicateurs avancés (tels que, l'effet bonanza, le sudden stop, la croissance excessive des crédits, Le ratio dette à court terme/réserve, le ratio masse monétaire/réserve de change, la volatilité de taux de change et de l'inflation, les frictions politiques,…) et les suivre en continu afin de se prémunir de risques pouvant saper la stabilité financière et l'économie, et renforcer la régulation financière. La Banque Centrale du Congo devrait mettre en œuvre une orientation macroprudentielle pour atteindre de manière saine la stabilité financière et réduire les effets de la volatilité des flux de capitaux.

# CONCLUSION

Cet article a pour objet l'analyse de la boucle rétroactive entre la volatilité des flux de capitaux et la stabilité financière. C'est-à-dire d'analyser les interactions entre la volatilité des flux de capitaux et la stabilité financière.

Le système financier étant le lieu de rencontre des flux de capitaux (égalité entre épargne et investissement), une volatilité des flux de capitaux peut détruire le bon fonctionnement et la robustesse du système financier, c'est-à-dire saper la stabilité financière. De même un système financier faible, peu réglementé et mal géré peut exacerber la volatilité des flux de capitaux et in fine saper la stabilité financière.

Il était question de vérifier sur *il existence d'une boucle rétroactive entre la volatilité des flux de capitaux et la stabilité financière en RDC, et d'évaluer les contributions* des *politiques macroéconomique et macroprudentielle dans l'atténuation des effets de la volatilité de flux de capitaux sur la stabilité financière et dans la prévention de l'instabilité financière*

Eu égards aux résultats, il s'est avéré d'une part, qu'il y a une boucle rétroactive entre la volatilité des flux de capitaux et la stabilité financière. Et d'autre part, que les politiques macroéconomiques et prudentielles ne peuvent pas atténuer la volatilité des flux de capitaux et prévenir une instabilité financière. Il s'avère que l'approche macroprudentielle a donné un résultat meilleur que la politique monétaire. La mise en œuvre d'un cadre macroprudentielle par la Banque centrale du Congo serait bénéfique dans l'atteinte de la stabilité financière et dans l'atténuation de la volatilité des flux de capitaux.

La contribution de cet article est de préparer le système financier congolais à être résilient face à la volatilité des flux de capitaux et à la prévention des risques qui peuvent causer une instabilité financière, tout en se dotant des indicateurs d'alerte avancés pour mieux faire de prévision et renforcer l'infrastructure financière pour une stabilité financière durable.

# BIBLIOGRAPHIE

# ANNEXES

**Test de stabilité**

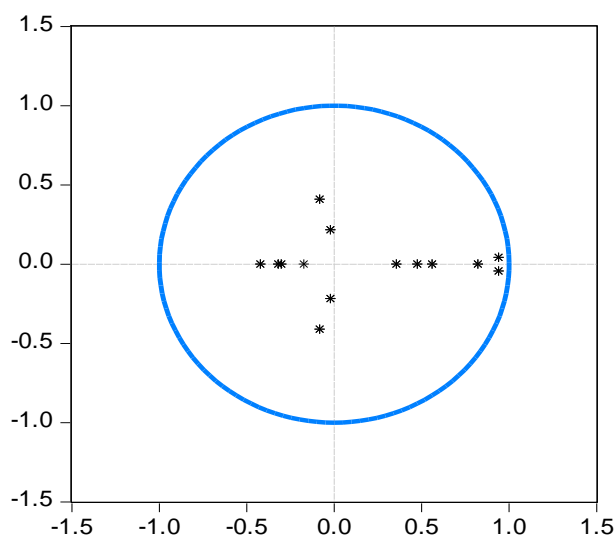

Inverse Roots of AR Characteristic Polynomial

La stabilité du modèle est une condition pertinente pour une bonne analyse de la politique macroéconomique et macroprudentielle. Le modèle est stable, étant donné que les racines sont dans le cercle.

**Test de causalité**

| Résultats des tests de causalité | | | |
|---|---|---|---|
| **Hypothèse nulle** | **Obs** | **F-Statistic** | **p-value** |
| DFNCPIB ne cause pas (au sens de granger) DFNPIB | 118 | 0.35622 | 0.5518 |
| DFNPIB ne cause pas (au sens de granger) DFNCPIB | | 0.10653 | 0.7447 |
| DFDET ne cause pas (au sens de granger) DFNPIB | 118 | 0.08802 | 0.7672 |
| DFNPIB ne cause pas (au sens de granger) DFDET | | 5.13474 | 0.0253 |
| DFCBI ne cause pas (au sens de granger) DFNPIB | 118 | 0.73473 | 0.3931 |
| DFNPIB ne cause pas (au sens de granger) DFCBI | | 0.05755 | 0.8108 |
| **DCPC ne cause pas (au sens de granger) DFNPIB** | **118** | **3.20908** | **0.0759*** |
| **DFNPIB ne cause pas (au sens de granger) DCPC** | | **4.05447** | **0.0464**** |
| DCOURSMPG ne cause pas (au sens de granger) DFNPIB | 118 | 0.00023 | 0.9880 |
| DFNPIB ne cause pas (au sens de granger) DCOURSMPG | | 0.00285 | 0.9575 |
| DPBPIB ne cause pas (au sens de granger) DFNPIB | 117 | 2.62457 | 0.1080 |
| DFNPIB ne cause pas (au sens de granger) DPBPIB | | 18.0678 | 4.E-05 |
| DFDET ne cause pas (au sens de granger) DFNCPIB | 118 | 0.45156 | 0.5029 |
| DFNCPIB ne cause pas (au sens de granger) DFDET | | 2.81749 | 0.0960 |
| DFCBI ne cause pas (au sens de granger) DFNCPIB | 118 | 3.9E-05 | 0.9950 |
| DFNCPIB ne cause pas (au sens de granger) DFCBI | | 0.17820 | 0.6737 |
| **DCPC ne cause pas (au sens de granger) DFNCPIB** | 118 | **3.84744** | **0.0522*** |
| **DFNCPIB ne cause pas (au sens de granger) Cause DCPC** | | **6.11379** | **0.0149*** |
| DCOURSMPG ne cause pas (au sens de granger) DFNCPIB | 118 | 4.90846 | 0.0287 |
| DFNCPIB ne cause pas (au sens de granger) DCOURSMPG | | 7.38111 | 0.0076 |
| DPBPIB ne cause pas (au sens de granger) DFNCPIB | 117 | 0.32199 | 0.5715 |
| DFNCPIB ne cause pas (au sens de granger) DPBPIB | | 0.72330 | 0.3968 |
| DFCBI ne cause pas (au sens de granger) DFDET | 118 | 1.78307 | 0.1844 |
| DFDET ne cause pas (au sens de granger) DFCBI | | 0.66482 | 0.4165 |
| DCPC ne cause pas (au sens de granger) DFDET | 118 | 0.30030 | 0.5848 |
| DFDET ne cause pas (au sens de granger) DCPC | | 0.26230 | 0.6095 |
| DCOURSMPG ne cause pas (au sens de granger) DFDET | 118 | 1.37122 | 0.2440 |
| DFDET ne cause pas (au sens de granger) DCOURSMPG | | 0.01058 | 0.9183 |
| DPBPIB ne cause pas (au sens de granger) DFDET | 117 | 1.28023 | 0.2602 |
| DFDET ne cause pas (au sens de granger) DPBPIB | | 1.07867 | 0.3012 |
| DCPC ne cause pas (au sens de granger) DFCBI | 118 | 0.07146 | 0.7897 |
| DFCBI ne cause pas (au sens de granger) DCPC | | 2.42221 | 0.1224 |
| DCOURSMPG ne cause pas (au sens de granger) DFCBI | 118 | 0.00048 | 0.9825 |
| DFCBI ne cause pas (au sens de granger) DCOURSMPG | | 0.01096 | 0.9168 |
| DPBPIB ne cause pas (au sens de granger) DFCBI | 117 | 1.99324 | 0.1607 |
| DFCBI ne cause pas (au sens de granger) DPBPIB | | 17.3061 | 6.E-05 |
| DCOURSMPG ne cause pas (au sens de granger) DCPC | 118 | 1.99878 | 0.1601 |
| DCPC ne cause pas (au sens de granger) DCOURSMPG | | 2.81782 | 0.0959 |
| DPBPIB ne cause pas (au sens de granger) DCPC | 117 | 0.00537 | 0.9417 |
| DCPC ne cause pas (au sens de granger) DPBPIB | | 1.25827 | 0.2643 |
| DPBPIB ne cause pas (au sens de granger) DCOURSMPG | 117 | 0.18816 | 0.6653 |
| DCOURSMPG ne cause pas (au sens de granger) DPBPIB | | 0.21965 | 0.6402 |

Nombre de retards inclus : 1

**Détermination du décalage optimal du processus VAR(1)**

| Lag | LogL | LR | FPE | **AIC** | **SC** | HQ |
|---|---|---|---|---|---|---|
| 0 | 7339.708 | NA | 7.27e+48 | 132.3731 | 132.5440 | 132.4424 |
| **1** | -6935.713 | 749.7559 | 1.21e+46* | **125.9768*** | **127.3438*** | 126.5313* |
| 2 | -6887.898 | 82.70678 | 1.25e+46 | 125.9982 | 128.5612 | 127.0379 |
| 3 | -6841.655 | 74.15605 | 1.35e+46 | 126.0478 | 129.8070 | 127.5728 |
| 4 | -6818.135 | 34.75028 | 2.25e+46 | 126.5069 | 131.4622 | 128.5171 |
| 5 | -6800.201 | 24.23436 | 4.27e+46 | 127.0667 | 133.2181 | 129.5621 |
| 6 | -6693.634 | 130.5690* | 1.72e+46 | 126.0294 | 133.3769 | 129.0101 |
| 7 | -6686.933 | 7.364541 | 4.46e+46 | 126.7916 | 135.3352 | 130.2575 |
| 8 | -6676.513 | 10.13855 | 1.17e+47 | 127.4867 | 137.2264 | 131.4378 |

\* indicates lag order selected by the criterion
LR: sequential modified LR test statistic (each test at 5% level)
FPE: Final prediction error
AIC: Akaike information criterion
SC: Schwarz information criterion
HQ: Hannan-Quinn information criterion

**Test d'autocorrélation**

| VAR Residual Serial Correlation LM Tests | | |
|---|---|---|
| Null Hypothesis: no serial correlation at lag order h | | |
| Lags | LM-Stat | Prob |
| 1 | 74.16596 | **0.1806** |
| 2 | 55.03953 | 0.7801 |
| 3 | 84.21320 | 0.0460 |
| 4 | 44.18706 | 0.9721 |

**Test de racine unitaire**

La stationnarité des séries a été testée à partir des tests de racines unitaires de Dickey-Fuller augmentés. Le cycle du crédit et le coussin de fonds propres contracyclique sont des composantes cycliques. Nous les considérons comme étant stationnaire.

| *Variables* | ADF | VCM au seuil de 5% | Ordre d'intégration | Décision |
|---|---|---|---|---|
| *FncPIB* | -4,5 | -3,4 | I(1) | Stationnaire |
| *FnPIB* | -4,6 | -3,5 | I(1) | Stationnaire |
| *Crocdf* | -11,1 | -3,4 | I(1) | Stationnaire |
| *Crousd* | -11,2 | -3,4 | I(1) | Stationnaire |
| *BBCC* | -3,6 | -3,4 | I(0) | Stationnaire |
| *Dnf* | -4,5 | -3,5 | I(1) | Stationnaire |